\documentclass[12pt]{article}
\usepackage{pic03}
\usepackage{hyperref}
\usepackage{url}
\usepackage{graphicx}

\begin{document}

\title{\bf A study of short-time periodic variation of the $^{8}$B solar neutrino flux at Super-Kamiokande}
\author{J.Yoo\\
{for the Super-Kamiokande collaboration}  \\
{\em  Department of Physics, Seoul National University,}\\
{\em San 51-1 Shinrimdong, Kwanakgu, Seoul, Korea}}
\maketitle

\baselineskip=14.5pt
\begin{abstract}
Super-Kamiokande(SK) is a real-time detector capable of measuring the exact time of solar neutrino events. This, combined with a relatively high yield of these events of roughly 15 per day, allows a search for short-time variations in the observed flux. Using all 1496 days of SK-I's solar data, we looked for periodic variations of the observed solar neutrino flux, and found no significant periodicity. 
\end{abstract}
\baselineskip=17pt
\section{Introduction}
	Super-Kamiokande (SK) is a water Cherenkov detector in Kamioka, Japan. Solar neutrino data were collected at SK from May 31st, 1996 to July 15th , 2001 yielding a total detector live time of 1,496 days (SK-I). The solar neutrino signal is extracted from the data using the angular deviation between the Sun and the reconstructed direction of events \cite{sk2001}. A total of $22,400 \pm 200$(stat.) solar neutrino interactions were observed in 22.5 ktons of fiducial volume. The relatively high yield of real-time events in SK, 15 events per day, allows a search for short-time periodic modulations in the observed neutrino fluxes. \par
\section{SK 10-day long sampled solar neutrino data}
\begin{figure}[h!]
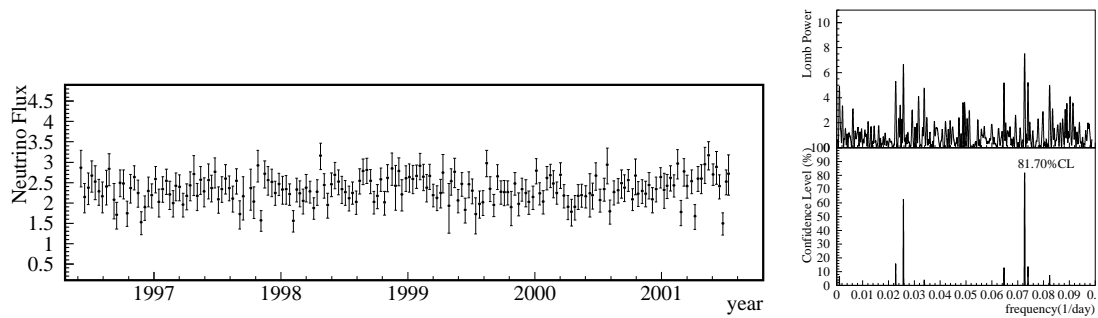

  \center{\includegraphics[width=10cm]{10daydata.epsi}{\hspace{0.5cm}}{\includegraphics[width=4cm]{power-cl-m2.epsi}}}
  \caption{ \it The left figure is measured solar neutrino fluxes of 10-day long samples. The right figure shows a Lomb periodogram of the SK 10-day long solar neutrino data samples. The Lomb power and its corresponding confidence level are given as a function of frequency.}\label{fig:10day}
\end{figure}

	The solar neutrino data, selected over 1,871 elapsed days from the beginning of data-taking, are divided into roughly 10-day long samples. The time period of each 10-day sample is chosen from consecutive 10-day periods from May 31st, 1996 through July 15, 2001. There are on and off periods of data-taking in the 10-day interval and thus the timing of each sample is calculated as a mean of the start and end times and corrected by SK livetime. Fig.\ref{fig:10day} (left) shows the measured solar neutrino fluxes of the 10-day samples. All given uncertainties are statistical. The 1/$R^2$ (squared average distance in A.U.) variation caused by the eccentricity of the Earth's orbit around the Sun is corrected for the measured solar neutrino fluxes.\par
\section{Search for Periodicity}
	The Lomb periodogram method, a spectral analysis for unevenly sampled data, is applied to search for possible periodicities in the measured 10-day long fluxes. The method finds periodicities based on maximum deviation of data relative to a constant in time. A detailed description of the method can be found in ref.\cite{nrc}.  The maximum power appears at frequency f=0.0726 day$^{-1}$ (or time period T=13.76 days) with Lomb power 7.51 corresponding to 81.70\% C.L. \par
 As a consistency check for the confidence level, 10,000 MC experiments are generated based on the observed timing information and the measured solar neutrino fluxes of the 10-day long data samples. The measured solar neutrino flux values are simulated according to a random Gaussian fluctuation. For making these null modulation samples the average of the measured fluxes ($2.33 \times 10^6$ cm$^{-2}$ s$^{-1}$) is taken as a Gaussian mean, and the standard deviation of the the measured flux ($0.32 \times 10^6$ cm$^{-2}$ s$^{-1}$) is taken as a Gaussian error \cite{horne}.
 The Lomb method is applied to each MC experiment to obtain a periodogram and the maximum power is selected. Out of 10,000 simulated experiments, 19.58\% have maximum powers larger than 7.51.  This demonstrates that the confidence level for the T=13.76 day period of SK data is consistent with that of no modulation.\par
\section{Sensitivity of Finding a Periodicity vs. Modulation Amplitude}
We have studied the sensitivity of the SK-I solar neutrino data to find if a true periodicity indeed exists. The details of this sensitivity study can be found in Ref.\cite{sk-mod}. The sensitivity of finding the correct period varies rapidly near the sampling time periods of 10 $\sim$ 20 days.  Based on the MC study, the Lomb method is expected to find the true periodicity of the SK-I data, if any, with an efficiency of 95\% at a modulation amplitude of 10\% of the averaged flux and a period of longer than 40 days in case of the 10-day long sample.\par
\section{Summary}
We have presented the measured solar neutrino fluxes of 10-day long samples using all 1,496 days of SK data. No significant periodicity was found in the SK solar neutrino data when a search was made to look for periodic modulations of the observed fluxes using the Lomb method. Based on a MC study, we have obtained the probability of finding a true periodicity in the SK data as a function of the modulation magnitude. The Lomb method should have found a periodic modulation in the SK-I solar neutrino data of 10-day long samples if the modulation period were longer than 40 days and its magnitude was larger than 10\% of the average measured neutrino fluxes.\par

\end{document}